\documentclass[runningheads]{llncs}

\usepackage{authblk}
\usepackage{cite}
\usepackage{array}
\usepackage{graphicx}
\usepackage{fix-cm}
\usepackage{times}
\usepackage{dirtytalk}
\usepackage{textcomp}
\usepackage{eurosym}
\usepackage{nomencl,etoolbox,ragged2e,siunitx}
\usepackage[utf8]{inputenc}
\usepackage[english]{babel}
\usepackage[table]{xcolor}
\usepackage{tcolorbox}
\usepackage{etoolbox}
\usepackage{tabu}
\usepackage{lastpage}
\usepackage[utf8]{inputenc}
\usepackage[table]{xcolor}
\usepackage{tikz}
\usepackage{fancyhdr}
\usepackage{amsmath}
\usepackage{kantlipsum}
\usepackage{lettrine}
\usepackage{enumitem}
\usepackage{rotating}
\usepackage{lscape}
\usepackage{tabularx}
\usepackage{todonotes}
\usepackage{dblfloatfix} 
\usepackage{amsmath}
\usepackage{multicol}
\usepackage{lipsum}
\usepackage{tikz}

\usepackage{float}

\usepackage{adjustbox}

\usepackage{graphicx}
\begin{document}
\title{ Keystroke Dynamics as Part of Lifelogging
\thanks{Alan Smeaton is partially supported by Science Foundation Ireland under Grant Number SFI/12/RC/2289\_P2.}
}
\titlerunning{Keystroke Dynamics and Lifelogging}
%
\author{Alan F. Smeaton\inst{1,2}\orcidID{0000-0003-1028-8389} \and
Naveen Garaga Krishnamurthy\inst{2} \and
Amruth Hebbasuru Suryanarayana\inst{2}}
%
\authorrunning{A.F. Smeaton et al.}
%
\institute{Insight Centre for Data Analytics, Dublin City University, Glasnevin, Dublin 9, Ireland \and
School of Computing, Dublin City University, Glasnevin, Dublin 9, Ireland\\
\email{Alan.Smeaton@DCU.ie}}
%
\maketitle              
\begin{abstract}
In this paper we present the case for including keystroke dynamics in lifelogging.  We describe how we have used a simple keystroke logging application called Loggerman, to create a dataset of longitudinal keystroke timing data spanning a period of more than 6 months for 4 participants.  We perform a detailed analysis of this data by examining the timing information associated with bigrams or pairs of adjacently-typed alphabetic characters.  We show how there is very little  day-on-day variation of the keystroke timing among the top-200 bigrams for some participants and for others there is a lot and this correlates with the amount of typing each would do on a daily basis. We explore how daily variations could correlate with sleep score from the previous night but find no significant relationship between the two.  Finally we describe the public release of this data as well including as a series of pointers for future work including correlating keystroke dynamics with mood and fatigue during the day.
\end{abstract}
%
%
%
\section{Introduction}

Lifelogging is the automatic gathering of digital records or logs about the activities, whereabouts and interactions of an ordinary person doing ordinary things as part of her/his ordinary day. Those records are gathered by the person, for the exclusive use of the person and not generally shared.  Lifelogs are a personal record which can be analysed either directly by the person collecting the data, or by others \cite{tuovinen2019remote}, in order to gain insights into long term behaviour and trends for wellness or behaviour change, as well as to support searching or browsing for specific information for the past.

Lifelogging as a practical activity has been around for many years and has matured as the technology to ambiently and passively capture daily activities has  evolved \cite{10.1561/1500000033}.
Technologies for capturing lifelog data are wide ranging and well-documented and can be broadly classified into wearable or on-body devices such as  wearable cameras, location trackers or physiological trackers for heart rate, respiration, etc. and off-body logging where sensors may form part of our environment like passive IR sensors for presence detection and contact sensors on doors and windows in the home. Off body logging would also include  using software such as measures for cumulative screentime viewing, productivity at work or online media consumption.
Whichever lifelog technolog(ies) may be used by a person, it is when these are combined and fused together that we get the best insights into the person as it is well accepted that so many aspects of our lives interact with, and depend on each other.

In a recent article by Meyer {\em et al.} \cite{10.1145/3373719}
the authors highlighted several current issues for longer term self-tracking. Some of these are technical issues including
incompleteness of data leading to data gaps, 
implicit tracking with secondary sources such as social networks and online services, and multiple interpretations of our data, beyond behaviour support.  There are also issues of self-tracking for  secondary users such as children or people with special needs with consequent
ethical, legal, and social implications \cite{jacquemard2014challenges}.

If we regard a lifelog collection as a multimedia or a multimodal artifact then it can take advantage of progress made in many other areas of multimedia analysis such as computer vision techniques to analyse images from wearable cameras \cite{WANG2018249}.  Progress in  areas like computer vision have depended upon the easy availability of large datasets on which new ideas can be evaluated and compared to previous work and initiatives like ImageNet have helped to catalyse these developments.  Yet when it comes to the general availability of lifelog collections, these are much rarer precisely because the data is personal.
In related research areas which use these same technologies like using wearable sensors for sleep or gait analysis in clinical settings \cite{johansson2018wearable} or using wearable cameras for measuring exposure to different food types \cite{signal2017children} then lifelog data collections do exist but in these cases the wearers are anonymised. In lifelogging it is the accumulation of data drawn together from across different sources and then fused together, that makes the lifelog and that does not reconcile well with the idea of  anonymisation.

In this paper we provide a brief review of past work on using keystroke information for user authentication, for identifying different stages of writing strategies, for measuring stress and emotion.  We advocate for greater use of keystroke dynamics in lifelogging
and we
describe a dataset of longitudinal keystroke and sleep data  gathered from 4 participants over a period of more than 6 months.
We describe an analysis of this dataset examining  its daily consistency over time both within and across participants
and we
address the anonymisation of this data by releasing it in aggregated form which allows within-participant and cross-participant comparisons.
The paper is organised as follows: in the next section we provide an overview of keystroke dynamics and then we describe the dataset we collected.  We then provide an analysis of this data showing its consistency over time and its comparison to sleep score data. Finally, in our concluding section we summarise the case for greater use of keystroke dynamics in lifelogging and point to future work.

\section{Keystroke Dynamics}
\label{sec:dynamics}

\begin{figure}[htb]
    \begin{center}
\includegraphics[width=0.8\textwidth]{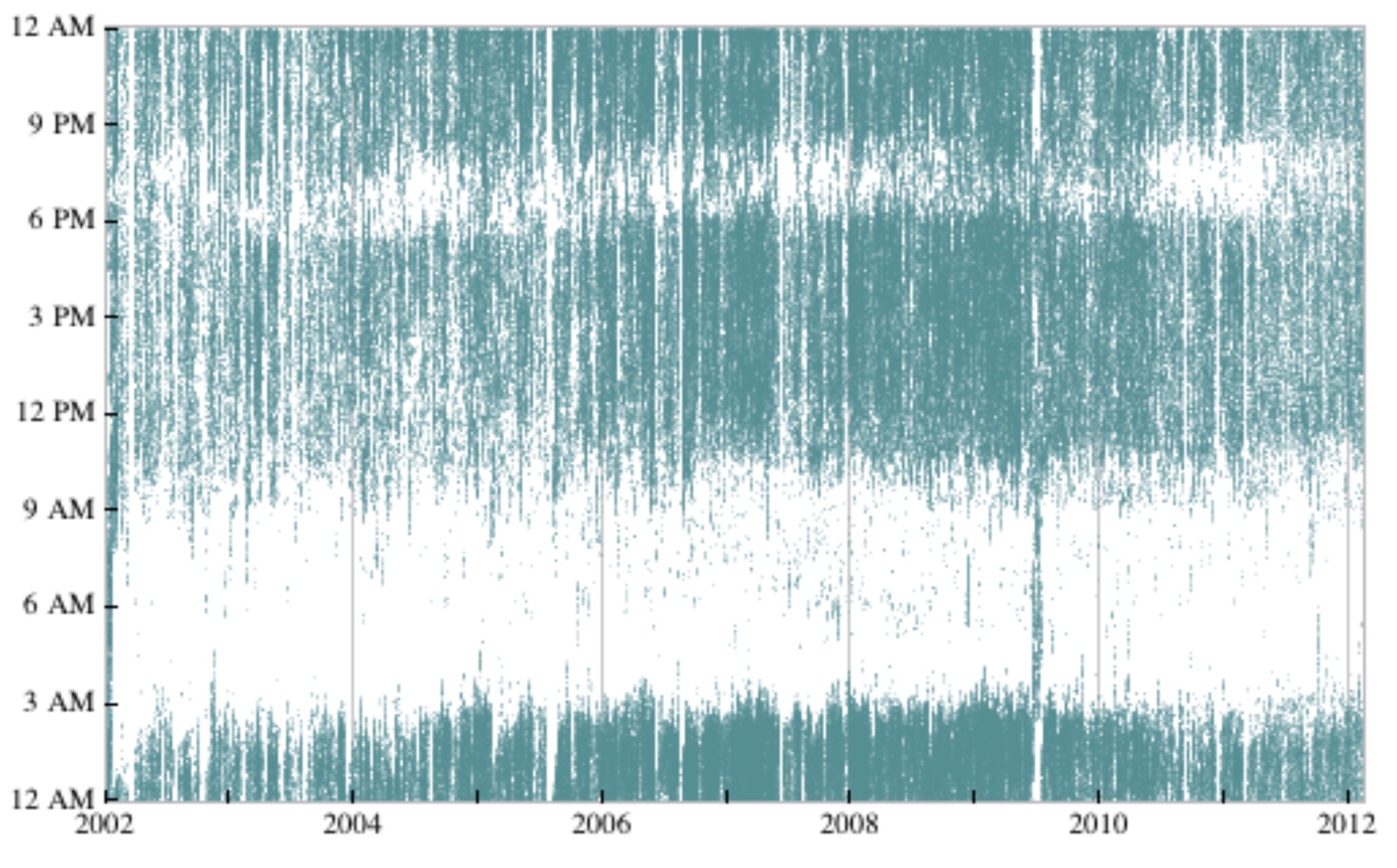}
\end{center}
    \label{fig:wolfram}
    \caption{Insights into timing for Stephen Wolfram's keystrokes over 10 years}
\end{figure}

In 2009 Stephen Wolfram reported that he had been using a keystroke logger that  collected a record of his every keystroke for the previous 22 years\footnote{\url{https://quantifiedself.com/blog/stephen-wolfram-keystroke-logg/}}.  This was in the form of the key pressed and the date and time of pressing.  By 2012 this had grown to be a record of 100 million keystrokes  \footnote{\url{https://writings.stephenwolfram.com/2012/03/the-personal-analytics-of-my-life/}} and from all this he was able to generate some interesting visualisations on usage and on his life, such as the one shown in Figure~\ref{fig:wolfram}.  This shows his rather interesting work patterns -- he basically works all day, and evening, stopping at about 3AM before resuming at about 10AM the following day with a break of a couple of hours, sometimes, in the evening for dinner. We can also see his various trips where he switched to local timezones such as his Summer of 2008 spent in Europe and there are other interesting facts  like that the average fraction of keys he types that are backspaces has consistently been about 7\%. 

\noindent  
While this kind of raw visualisation and analysis may be interesting, it is only when we add detailed timing information, like recording keystroke times to the nearest millisecond so that we can look at inter-keystroke times, i.e. the time needed to type two or more adjacent characters, that we can get  different kinds of insights into  participants.

The original application for keystroke dynamics with accurate timing information was as a form of  user authentication and work in this area goes back over 4 decades, from 1980  onwards and with regular re-visits to the topic \cite{gaines1980authentication,joyce1990identity,bergadano2002user}. The security application for keystroke dynamics is based on the premise that each of us have our own unique timing information as we interact with GUIs and that includes the timings of our keystrokes, our mouse movements, and our mouse clicks \cite{hinbarji2017user}. 
An advantage of using keystroke dynamics for security and authentication would be that we would never need to remember passwords, and passwords could never be hacked because they would be replaced by our keystroke dynamics.  However the way authentication for access to our computer systems has developed over the last half-century is that they present as tests to be overcome at the point of entry, similar to the way a passport is used at an airport. Keystroke dynamics take some time for baseline timing patterns to emerge and become established  and thus they are not useful for authentication at point of entry, which is why we do not see it in common use today.

Keystroke logging has had other more successful applications including  identifying different kinds of author writing strategies and 
understanding cognitive processes  \cite{leijten2013keystroke}. The premise here is that we establish a baseline for our keystroke timing information gathered over a long period and at any given period during the day we can compare the current dynamics with the baseline to see if we are typing faster, or slower, perhaps indicating that we are in full creative flow or that we are pondering our thoughts as we write. This also exploits pause location as we type and whether pauses occur between words, between sentences or even between paragraphs and what insights into the author's thinking can be gleaned from such pauses \cite{leijten2019analysing}.

Keystroke timing information has also been used for measuring stress \cite{vizer2009detecting} where the authors found that it is possible to classify cognitive and physical stress conditions relative to non-stress conditions based on keystroke and text features.  It has also been used for   emotion detection where \cite{kolakowska2013review} provides a review of almost a dozen published papers addressing this specific topic, and that review was from 2013.

What previous work shows  is that keystroke dynamics can provide untapped insights into our behaviour in a way which is non-intrusive, requires no investment in hardware, uses up a miniscule amount of computer resources yet this is a  data source that we have largely ignored to date.
In this paper we argue for keystroke logging as a data source for lifelogging and we illustrate our case using keystroke information collected from 4 participants over more than 6 months.

\section{Collecting Keystroke Data}

\begin{figure}[htb]
    \centering
    \includegraphics[height=0.3cm]{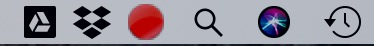}
    \hspace{0.1\textwidth}
    \includegraphics[height=0.3cm]{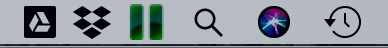}
    \caption{Loggerman status  indicator on  a Macbook, third icon from left, showing Loggerman recording (left side) and Loggerman paused (right side)}
    \label{fig:display}
\end{figure}

For collecting keystroke dynamics we used Loggerman \cite{hinbarji2016loggerman} a comprehensive logging tool which can capture many aspects of our computer usage including  keyboard, mouse and interface actions. This information is gathered ambiently and stored on the local computer.
For keystrokes, Loggerman can record  complete words typed by the participant where words are separated by whitespace.  When the participant is typing a password, recording is automatically disabled.  Once installed, Loggerman can simply record information to log files and uses a minuscule amount of CPU time.  The status of Loggerman, i.e. whether it is recording or if it has been paused by the participant, appears on the computer GUI as an icon on the menu bar.  Figure~\ref{fig:display} shows two versions of the menu bar on an Apple Macbook with Loggerman enabled  and with Loggerman paused by the participant.

\noindent  
From an examination of Loggerman files across several participants we  see that participants regularly make use of autocomplete, they make typing errors and then use backspace or they re-position their cursor with arrow keys to change a previously mis-typed word or to fix a spelling error. Thus the amount of fully-typed and correctly-typed words in Loggerman's word file is much less than we anticipated. The keystroke dynamics associated with such instances of  cursor navigation and re-positioning will not be reflective of the ideal creative flow  that we would like when we type and thus the keystroke timing information for the overall logging period will have been ``polluted'' by this necessity of  correcting typing errors or of re-phrasing.  That is unfortunate, and some participants may have more of this than others and even a given participants may have periods of more or less of the ``flow'' experience, as we discussed earlier when presenting keystroke logging for investigating writing strategies.

To illustrate the potential of keyboard dynamics in lifelogging we gathered information using Loggerman from 4 participants covering 1 January 2020 to 24 July 2020 (206 days) and we present an analysis of data from  those subjects.  Table~\ref{tab:particip} shows the amount of data generated. The number of days logged varies per participant because participants might disable logging and forget to resume it, and we also see an almost tenfold variation in the number of keystrokes typed on average per day between participants 1 and 4, with 2 and 3 in between.

\begin{table}[!ht]
    \centering
        \caption{Keystroke information for our 4 participants
    \label{tab:particip}}
    \begin{tabular}{c|c|c|c}
\hline 
&Total Keystrokes &	No of logged days&	Average Keystrokes/day \\ \hline \hline
Participant 1 &	2,174,539	&219	&9,929 \\
Participant 2 &	1,147,285	&189	&6,070 \\
Participant 3 &	802,515	&227	&3,535 \\
Participant 4 &	92,967	&90&	1,033 \\ \hline
    \end{tabular}
\end{table}

%
%
%
\vspace{-0.5cm}
\noindent 
When we analysed the log files from across participants we found that many of the most frequently used characters are special characters like punctuation marks and numbers as well as keys for cursor navigation. For the purpose of our timing analysis we will not consider these special characters since they are not part of normal typing flow and so we consider only  alphabetic characters A to Z. This reduces the number of keystrokes by almost half, so for participant 1 the total of 2,174,539 keystrokes reduces to 1,220,850 typed characters.
For timing purposes we treat uppercase and lowercase as equal. A
rationale for doing this because it reduces the number of possible 2-character strings (bigrams) we work with to $26 \times 26 = 676$ possible combinations.

As mentioned earlier, a lifelog's usefulness increases when there are multiple sources of logged data gathered by the participant.  Logging data on mood, emotion, stress or writing style at a given time were beyond the scope of this work which focuses on keystroke dynamics only however in addition to keystroke logging we also gathered information on participants' sleep.
There are a range of sleep tracking devices available off-the-shelf \cite{SHELGIKAR2016732} and we used the \={O}ura ring \cite{koskimaki2018we}.  This is a smart ring with in-built infrared LEDs, NTC temperature sensors, an accelerometer, and a gyroscope all wrapped into a ring form factor which gathers data for up to 7 days between charges.  During sleep it measures heart rate including heart rate variability, body temperature, and movement from which it can calculate  respiration rate.
From its raw data it computes a daily activity score, average METs, walking equivalent, a readiness score and for sleep it records bedtime, awake time, sleep efficiency, total sleep and several other metrics, including an overall sleep score.  From among all these options we use the overall score, a measure in the range 0 to 100 calculated using a proprietary algorithm which  is a function of total sleep, sleep efficiency, restfulness, REM and deep sleep, latency and timing. \={O}ura's interpretation of the sleep score is that if it is 85 or higher that corresponds to an excellent night of sleep, 70-84 is a good night of sleep while under 70 means the participant should pay attention to their sleep.

Our participants used a sleep logger for most of the 206 days of logging and for nights when the logger was not used we used a simple data imputation  to fill the gap. 

\section{Data Analysis}

In 2013 Peter Norvig  published the results of his computation of letter, word and n-gram frequencies drawn from the Google Books collection of 743,842,922,321 word occurrences in the English language\footnote{\url{http://norvig.com/mayzner.html}}.
In this he found the top 5  most frequently occurring bigrams in English are
TH, HE,  IN,   ER and   AN, though some of the possible 676 bigrams will never or almost never appear, such as JT, QW or ZB.  In our first analysis we focus on participant 1 as s/he gathered the largest volume of log data.
From among the 369,467 individual words typed  over 206 days, the top 10 most frequently occurring bigrams for all 4 participants are  shown in Table~\ref{tab:my_label}, along with the top 10 as found from Norvig's analysis.

\begin{table}
\centering
    \caption{10 most frequently used bigrams for participants\label{tab:my_label}}
    \begin{tabular}{|c|cccccccccc|}
    \hline
    Participant&    \multicolumn{10}{c|}{Bigram}  \\
    \hline \hline 
1 &TH  &IN  &HE &AN &RE &ER &ON &AT &ES &ND \\ 
2 &SS  &IN  &TH	&RE	&AT	&CV	&ES	&ER	&HE	&ON   \\
3 &IN  &RE  &AN	&AT	&ES	&ER	&SS	&CV	&TI	&ON   \\
4 &IN  &TH  &AN	&RE	&ER	&HE	&AT	&ON	&ES	&TE   \\ \hline \hline
Norvig's analysis& TH  &HE  &IN  &ER  &AN  &RE  &ON & AT & EN &ND \\
    \hline
    \end{tabular}
\end{table}

\begin{figure}[!t]
    \begin{center} 
        \includegraphics[width=\textwidth]{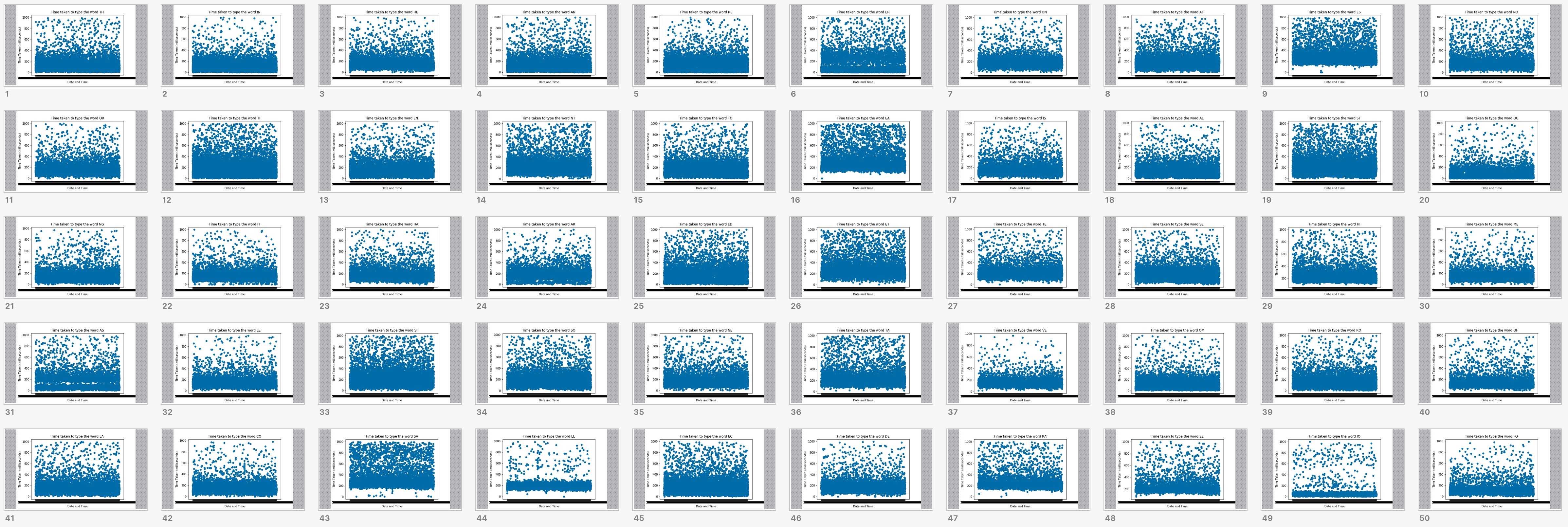}
        \includegraphics[width=\textwidth]{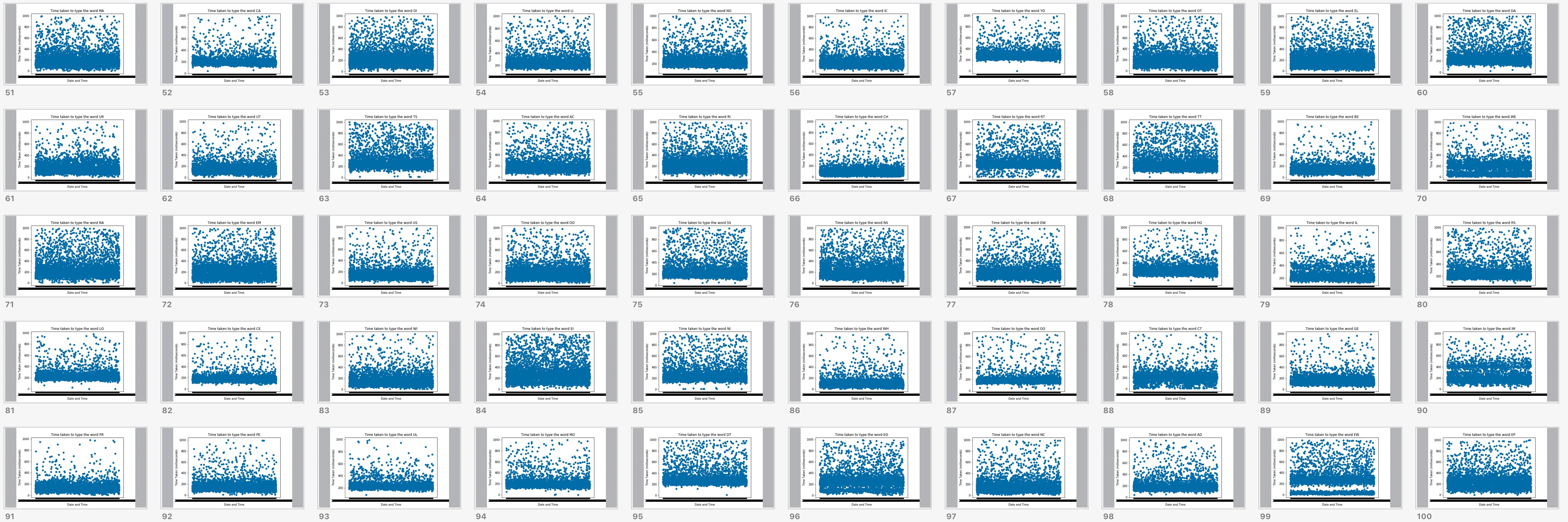}
        \includegraphics[width=\textwidth]{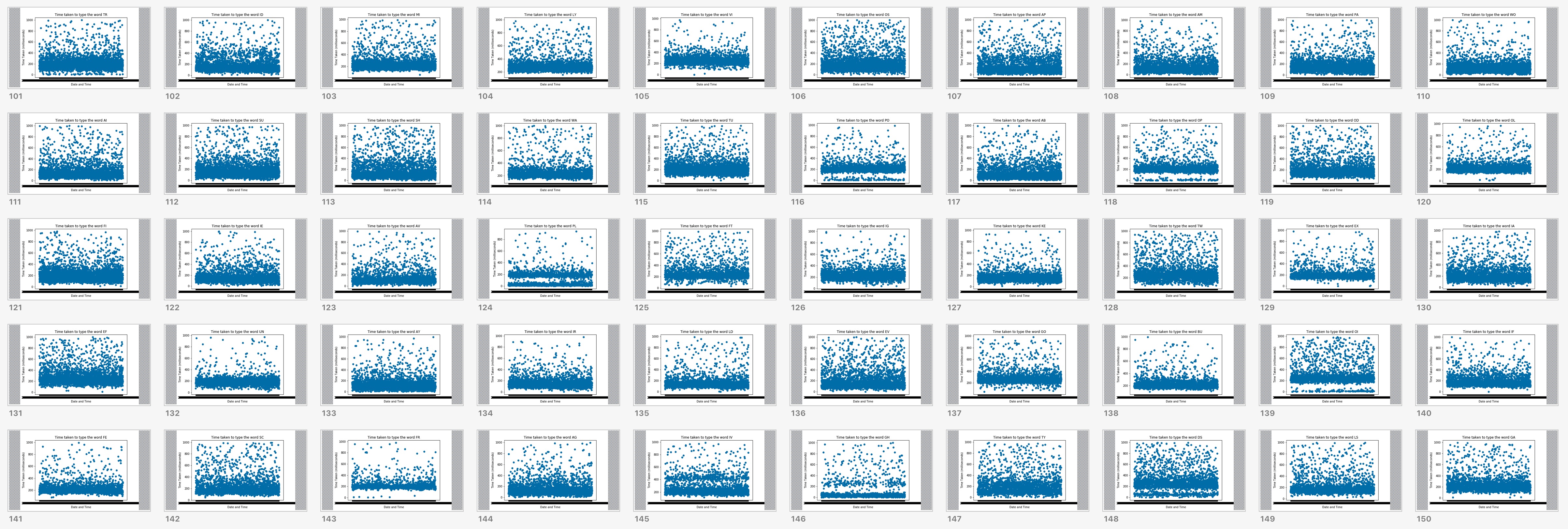}
        \includegraphics[width=\textwidth]{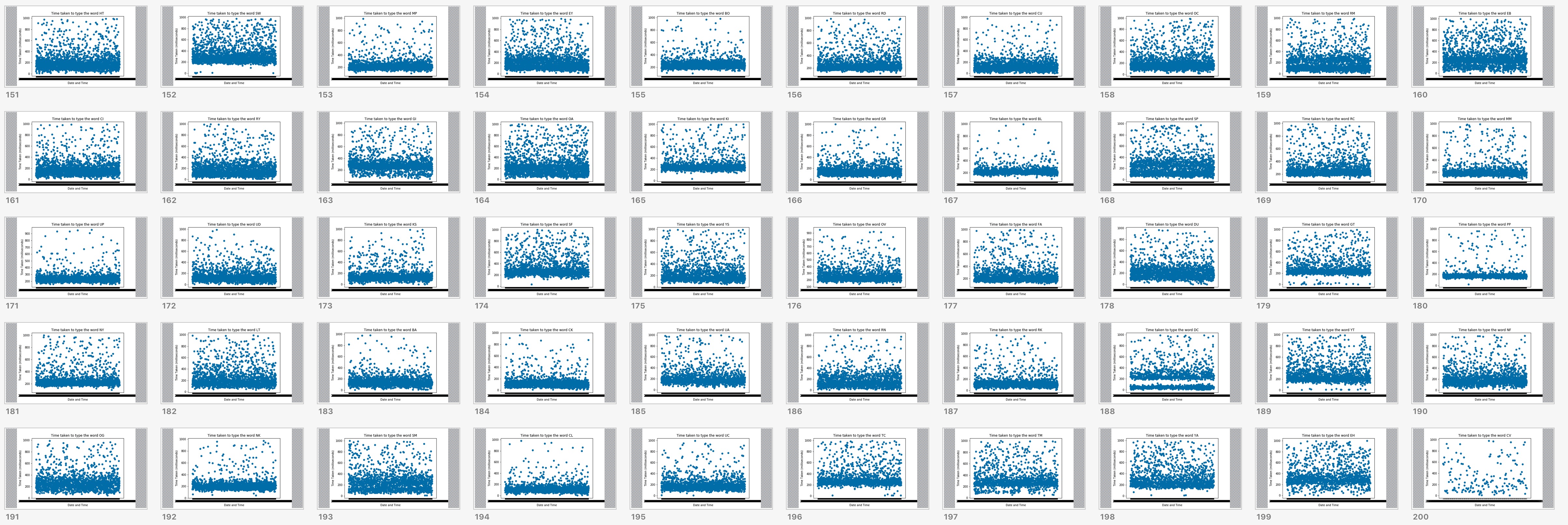}
    \end{center}
    
    \caption{Timing intervals over 206 days for participant 1 for each of the top-200 bigrams ranked by mean overall speed\label{fig:all-freqs}}
\end{figure}

\vspace{-0.5cm}
\noindent
This shows us that there is very little overlap among the top 10 actual bigrams typed by different participants but we are not interested in the actual bigrams typed but in the timing of that typing.
The distributions of timing information for each of the 200 most frequent bigrams over the 206 day logging period for participant 1 is shown as Figure~\ref{fig:all-freqs}. These individual graphs are too small to see any detail, but it is clear that the actual timing patterns for bigrams  vary quite a lot among these top 200.  For these graphs and the subsequent ones in this paper, we do not include inter-character timing gaps greater than 1,000ms and the graphs show the time taken for instances of each bigram plotted left to right from 1 January to 24 July.  

\begin{figure}[!ht]
    \begin{center}
\includegraphics[width=0.7\textwidth]{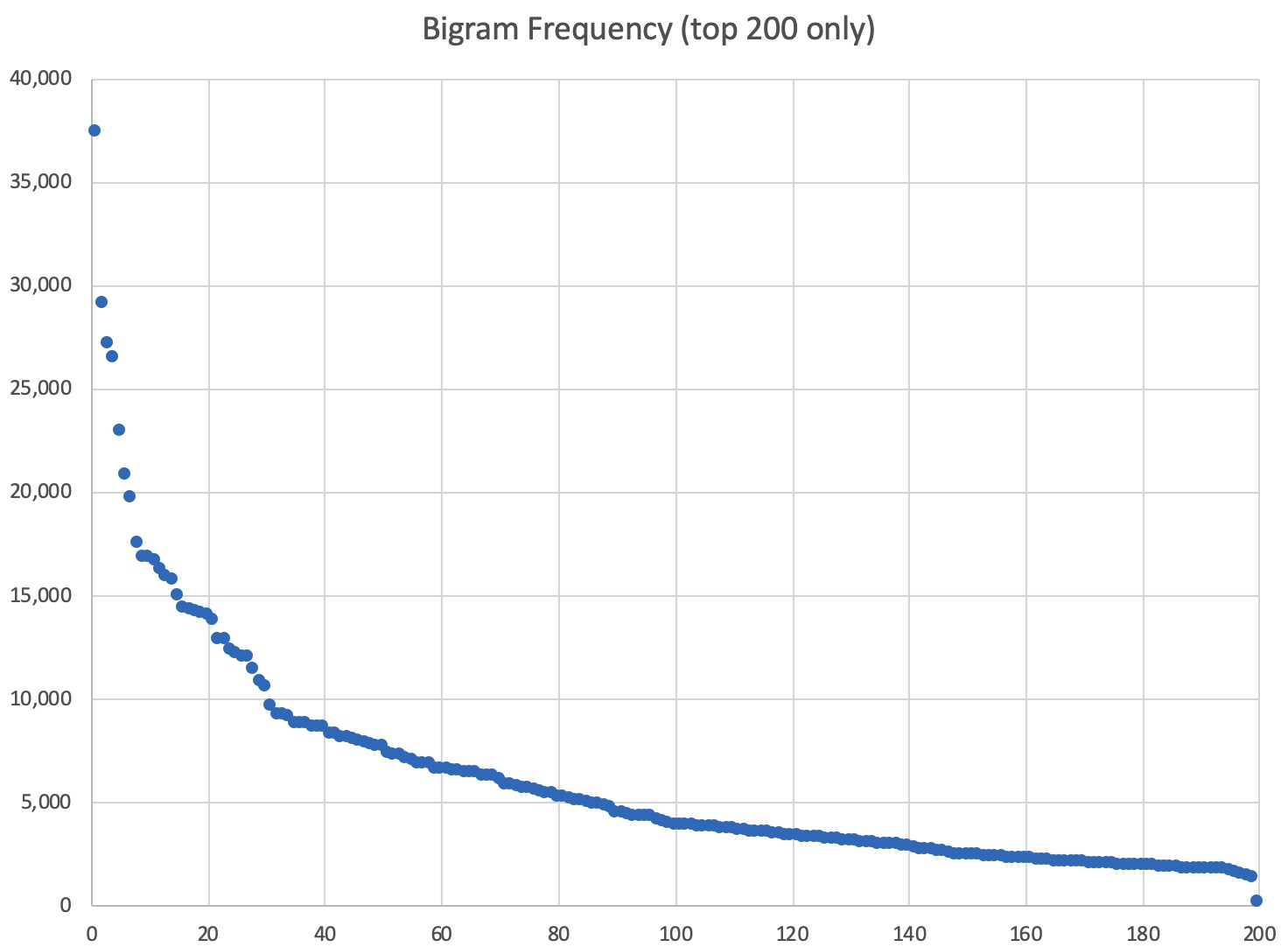}
\end{center}
    \caption{Frequency of occurrence for  most frequent bigrams (top 200 only) for participant 1
    \label{fig:bigram-freq}}
\end{figure}

\begin{figure}[!ht]
    \begin{center}
\includegraphics[width=0.7\textwidth]{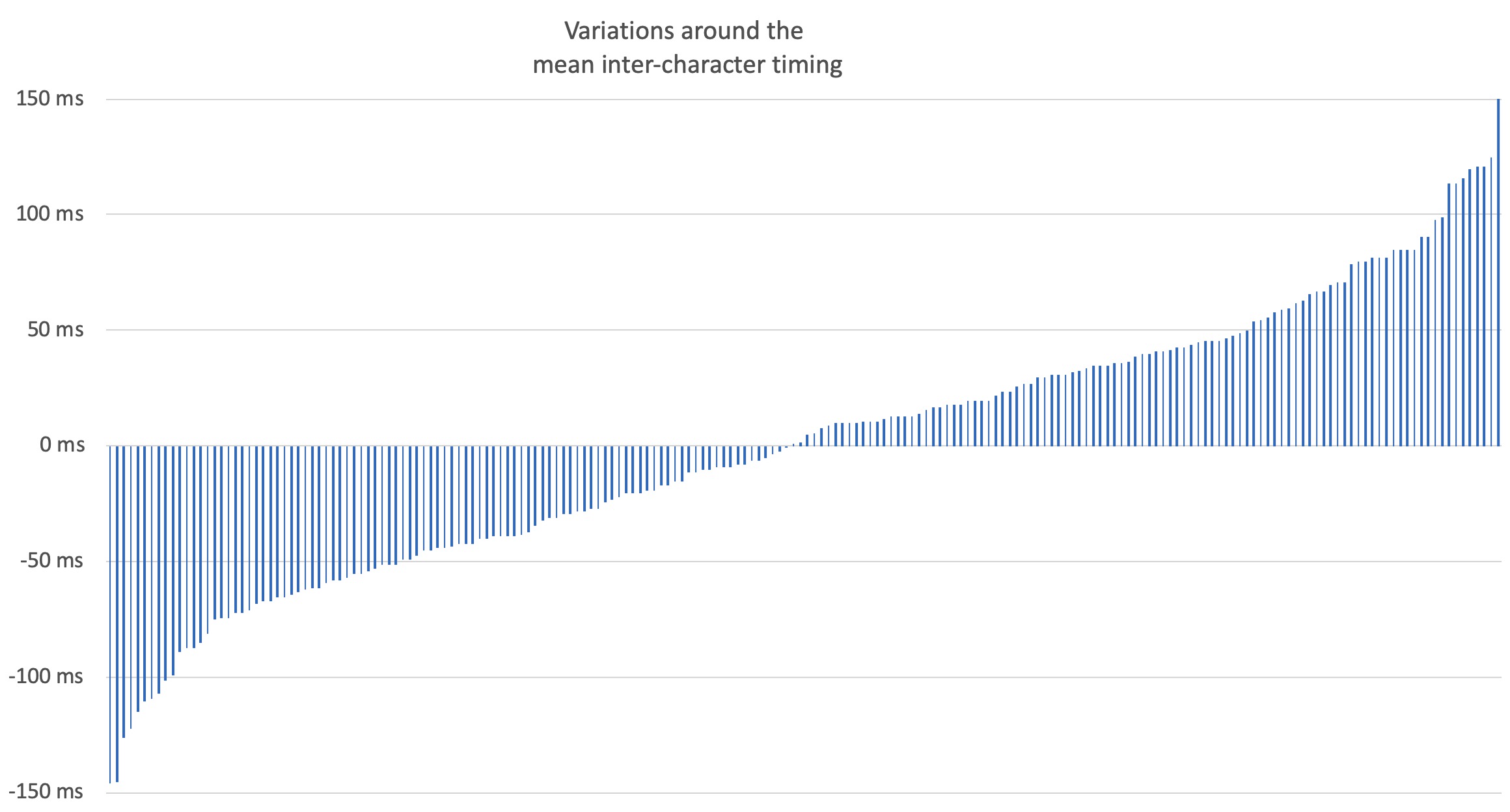}
\end{center}
    \caption{Spread or variability of meaning timing information for participant 1
    \label{fig:variab}}
\end{figure}

In Figure~\ref{fig:bigram-freq}  shows the  frequencies of occurrence for those top-200 most frequently used bigrams from participant 1 highlighting  that there are  a small number of very frequently occurring bigrams and then it tails off, in a Zipfian-like manner. This pattern is repeated for our other participants. When we look at how mean typing speeds for these top-200 bigrams from  across the 206 days vary compared to the overall mean for participant 1, which is 204ms,  there are a very small number of bigrams up to 150ms faster than the average and a small number of bigrams up to 150ms slower than the average. Most of the rest of these, approx 80\%, are between 75ms faster and 75ms slower than the average. Thus a clustering of approximately 80\% of bigram mean timings are within an overall range of only 150ms as shown in Figure~\ref{fig:variab}.

\begin{figure}[!ht]
    \begin{center}
\includegraphics[width=0.32\textwidth]{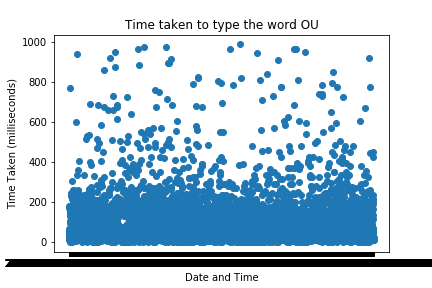}
\includegraphics[width=0.32\textwidth]{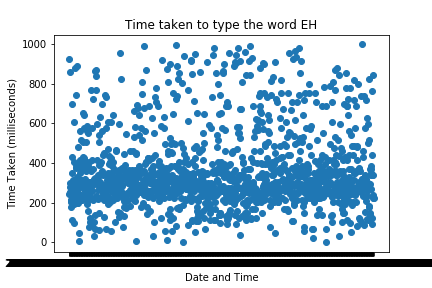}
\includegraphics[width=0.32\textwidth]{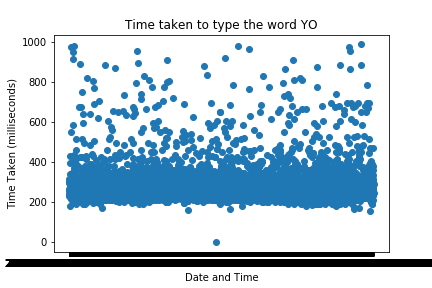}
\end{center}

    \caption{Timing information for fastest mean OU (ranked $1^{st}$), slowest mean from top 200 EH (ranked $200^{th}$), and slowest baseline (never fast) YO  (ranked $57^{th}$ for participant 1)
        \label{fig:speeds}}
\end{figure}

The mean and standard deviations for some  bigram timings for participant 1 are shown in Figure~\ref{fig:speeds}.  The fastest average of the 676 bigrams is OU with a mean time of 58ms but with very large standard deviation of	83.3 while  the slowest from among the top 200 bigrams is EH with a mean time of 358ms and standard deviation of	191. The bigram YO (ranked 179th most frequently occurring) with a mean time of 283ms and standard deviation of	81 has an interesting characteristic of never, ever being faster than about 200ms.  This can only be explained as a quirky characteristic of the keyboard typing of  participant 1 and we compare her/him to the others later.

For participant 1 we found that some of the bigrams (XV and VV) have an average timing which is over 500ms slower than the overall  average, indicating that this participant has  trouble finding the XV and VV character combinations, while some other bigrams like EI and IN are 162ms and 151ms faster than the average.  This might be due to the fingers usually used by her/him to type these particular character combinations.  We would expect that when using the middle and index fingers consecutively on keys that are adjacent and on the same row of the keyboard this would be faster to type than, say, using the little and then the index finger on two keys that are on a lower and then a higher row of the keyboard. On checking with the participant as to which fingers s/he uses to type the fastest of the bigrams we find that it is indeed the middle and index fingers for keys which are on the same row of the keyboard.  Thus some of the timing characteristics may be explained by keyboard layout and the particular fingering that a subject will use for pressing different keys.

\begin{figure}[!ht]
    \begin{center}
\includegraphics[width=0.32\textwidth]{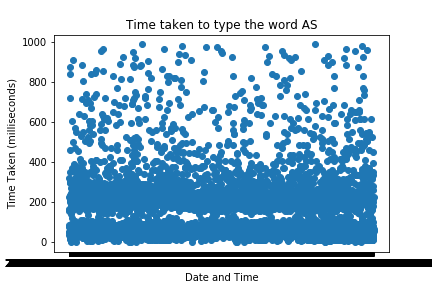}
\includegraphics[width=0.32\textwidth]{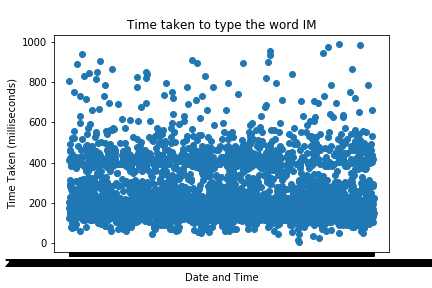}
\includegraphics[width=0.32\textwidth]{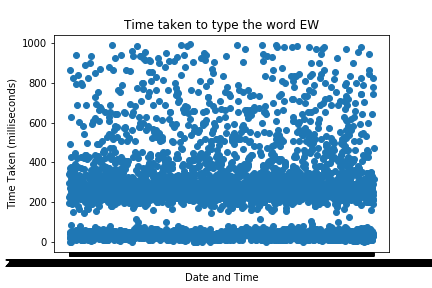}
\\
\includegraphics[width=0.32\textwidth]{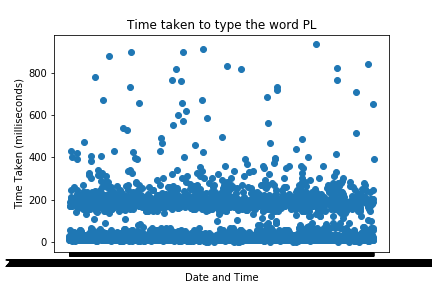}
\includegraphics[width=0.32\textwidth]{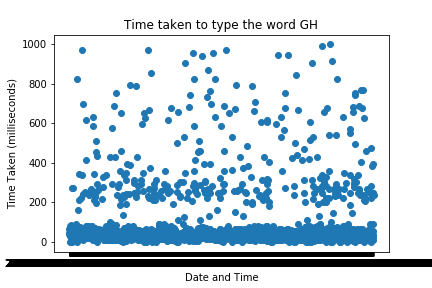}
\includegraphics[width=0.32\textwidth]{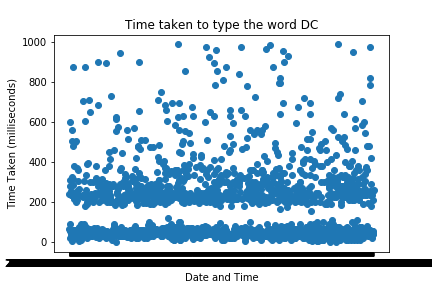}
\end{center}
    \caption{Timing  for AS, IM, EW, PL, GH  and DC for participant 1 showing banding effect
        \label{fig:banding}}
\end{figure}

We also discovered a strange banding effect for this participant's timing information which is shown in Figure~\ref{fig:banding}.  For bigrams AS (ranked $31^{th}$), IM ($90^{th}$), EW ($99^{th}$), PL ($124^{th}$), GH ($146^{th}$) and DC ($188^{th}$) there is a lower (faster) band of rapidly typed characters spanning right across the 207 day logging period with a gap in timing  before the more regular characteristic pattern of dense occurrences leading to more scattered occurrences as we approach 1,000ms.  Our only explanation for this is that it is to do with the rapid typing of a regularly used word among this participant's wordlist but this needs to be investigated further

\noindent 
We now look at  consistency of  bigram timing characteristics for participants across each  day of their logging period.
If we take the top-200 most frequent bigrams and rank order them by their mean speed for each of their logging days and then correlate the bigram rankings for each day, the average pairwise correlation for participant 1 is 0.6262.  The reader is reminded that these correlations are the average across $206 \times 206$ individual daily correlations of the top 200 bigrams so these are day-on-day variations in rank ordering of bigram mean typing speeds.   If we reduce the number of bigrams considered to the top 50, 25 and then 10  we see this correlation across the days increase, as shown in the first column of Table~\ref{tab:bigram-correl}, though when we get to the top 5 bigrams the correlation drops to 0.7937 which is explained by the fact that the top-5 mean fastest  bigrams may vary from day to day. This is highlighted in our comparison to Peter Norvig's analysis at the start of this section.  Table~\ref{tab:bigram-correl} also shows the same analysis applied to the other 3 participants 
and from this we see the others ave a much lower correlation among the bigrams they have typed fastest.  Also included in Table~\ref{tab:bigram-correl} and taken from Table~\ref{tab:particip} earlier, is the average number of keystrokes typed by each participant per day and we can clearly see that the more a participant types, the greater the consistency of their typing speeds.  Putting this  in other words, no matter what day it is,  participant 1 has almost the same ordering of her/his ranked bigram timings  whereas for the others, that ordering will vary more, probably because they type less than participant 1.

\begin{table}[!ht]
\centering
    \caption{Correlation among top bigrams ranked by frequency for logging period\label{tab:bigram-correl}}
    \begin{tabular}{|c||c|c|c|c|} \hline
Number of bigrams	&Participant 1&	 Participant 2	&Participant 3	&Participant 4\\ \hline \hline 
Top 5	&0.793	&0.515	&0.371	&0.251\\
Top 10	&0.921	&0.456	&0.437	&0.311\\
Top 25	&0.867	&0.233	&0.239	&0.357\\
Top 50	&0.829	&0.16	&0.234	&0.318\\
Top 200	&0.626	&0.214	&0.238	&0.302\\ \hline \hline 
Avg. keystrokes per day	&9,929 &6,070 &3,535 &	1,033 \\ \hline
\end{tabular}
\end{table}

\vspace{-0.5cm}
\noindent 
As mentioned previously, a lifelog becomes most useful when there are multiple sources of data which are then cross-referenced to gain insights into the participant's life.
We saw  in section~\ref{sec:dynamics} how keystroke dynamics has been used for measuring stress \cite{kolakowska2013review}, emotion   \cite{kolakowska2013review} and even our level of creative flow when writing \cite{leijten2019analysing}.
Using the sleep score data gathered by the \={O}ura ring, we explored whether sleep score correlates with bigram timing for any or all of our top-200 bigrams.  Mean daily timing data for the bigram TH had a +0.209 correlation with sleep score from the previous night while mean daily timing data for CV had a correlation of -0.18.  The average of these bigram correlations with sleep score was +0.014 which 

%
%
leads us to conclude that there is no correlation between daily typing speed of any bigrams and  sleep score, meaning, in turn, that  participant 1 is  consistent in typing speed, even when tired from poor sleep the previous night. When we applied this to other participants we found the same result. Perhaps if we explored windowing sleep score as a moving average over a number of days, or used other metrics to measure fatigue then that might correlate with timing information. 

Another possibility is that
bigram timing information might vary during the day, differing from morning to evening as fatigue effects might alternate with bursts of energy or bursts of enthusiasm or as the participant's level of stress or emotion or creativity might vary but that would require a more comprehensive lifelog. This goes back to the point we made in the introduction about the best kind of lifelog being a multimodal artifact, a fusion across multiple, diverse information sources.  The exercise reported in this paper has served to illustrate the possibilities that keystroke dynamics have as one of those information sources.

\section{Conclusions}

In this paper we present  the case for greater use of keystroke dynamics in lifelogging.  We report on several previous applications for keystroke dynamics and we describe how we used a tool called Loggerman to gather keystroke data for 4 participants over more then 6 months.  We are particularly interested in the timing information associated with keystrokes ad we showed how timing information between bigram keystrokes can vary for the same participant across different days. We also showed how the relative speeds with which bigrams are typed varies hugely for the same participant and also across different participants, showing how useful keystroke dynamics can be for security and authentication applications.

Keystroke dynamics has been shown to correlate with stress, fatigue and writing style and in this preliminary analysis we explored whether keystroke timing was correlated with fatigue, as measured by sleep score from the previous night. Unfortunately we found no correlation between these suggesting that a simple sleep score is insufficient to measure participant fatigue and that we need more fine-grained measures which would allow levels of fatigue which vary throughout the day, to be measured.

For future work there are a range of ways in which data from keystroke dynamics could be used as part of a lifelog, especially to gain insights into the more complex cognitive processes in which we engage each day.  Keystroke timing information has been shown to reveal writing strategies on the conventional keyboard/screen/mouse setup but it would be interesting to explore keystroke dynamics on mobile devices and see how that correlates with stress, cognitive load from multi-tasking, fatigue and distraction.

\noindent 
{\bf Data Availability} Keystroke usage data including bigram ranking, timing information and sleep information for our participants will be made available with a published version of this paper allowing other researchers to see  how keystroke timing information  can be  analysed and correlated with other lifelog data.
%
%
%

\bibliographystyle{splncs04}
\bibliography{bibfile}

\begin{thebibliography}{10}
\providecommand{\url}[1]{\texttt{#1}}
\providecommand{\urlprefix}{URL }
\providecommand{\doi}[1]{https://doi.org/#1}

\bibitem{bergadano2002user}
Bergadano, F., Gunetti, D., Picardi, C.: User authentication through keystroke
  dynamics. ACM Transactions on Information and System Security (TISSEC)
  \textbf{5}(4),  367--397 (2002)

\bibitem{gaines1980authentication}
Gaines, R.S., Lisowski, W., Press, S.J., Shapiro, N.: Authentication by
  keystroke timing: Some preliminary results. Tech. rep., Rand Corp Santa
  Monica CA (1980)

\bibitem{10.1561/1500000033}
Gurrin, C., Smeaton, A.F., Doherty, A.R.: Lifelogging: Personal big data.
  Foundations and Trends in Information Retrieval  \textbf{8}(1),  1–125 (Jun
  2014). \doi{10.1561/1500000033}

\bibitem{hinbarji2017user}
Hinbarji, Z., Albatal, R., Gurrin, C.: User identification by observing
  interactions with {GUI}s. In: International Conference on Multimedia
  Modeling. pp. 540--549. Springer (2017)

\bibitem{hinbarji2016loggerman}
Hinbarji, Z., Albatal, R., O’Connor, N., Gurrin, C.: Loggerman, a
  comprehensive logging and visualization tool to capture computer usage. In:
  International Conference on Multimedia Modeling. pp. 342--347. Springer
  (2016)

\bibitem{jacquemard2014challenges}
Jacquemard, T., Novitzky, P., O’Brolch{\'a}in, F., Smeaton, A.F., Gordijn,
  B.: Challenges and opportunities of lifelog technologies: A literature review
  and critical analysis. Science and engineering ethics  \textbf{20}(2),
  379--409 (2014)

\bibitem{johansson2018wearable}
Johansson, D., Malmgren, K., Murphy, M.A.: Wearable sensors for clinical
  applications in epilepsy, {P}arkinson’s disease, and stroke: a
  mixed-methods systematic review. Journal of neurology  \textbf{265}(8),
  1740--1752 (2018)

\bibitem{joyce1990identity}
Joyce, R., Gupta, G.: Identity authentication based on keystroke latencies.
  Communications of the ACM  \textbf{33}(2),  168--176 (1990)

\bibitem{kolakowska2013review}
Ko{\l}akowska, A.: A review of emotion recognition methods based on keystroke
  dynamics and mouse movements. In: 2013 6th International Conference on Human
  System Interactions (HSI). pp. 548--555. IEEE (2013)

\bibitem{koskimaki2018we}
Koskim{\"a}ki, H., Kinnunen, H., Kurppa, T., R{\"o}ning, J.: How do we sleep: a
  case study of sleep duration and quality using data from \={O}ura ring. In:
  Proceedings of the 2018 ACM International Joint Conference and 2018
  International Symposium on Pervasive and Ubiquitous Computing and Wearable
  Computers. pp. 714--717 (2018)

\bibitem{leijten2019analysing}
Leijten, M., Van~Horenbeeck, E., Van~Waes, L.: Analysing keystroke logging data
  from a linguistic perspective. In: Observing Writing, pp. 71--95. Brill
  (2019)

\bibitem{leijten2013keystroke}
Leijten, M., Van~Waes, L.: Keystroke logging in writing research: Using
  inputlog to analyze and visualize writing processes. Written Communication
  \textbf{30}(3),  358--392 (2013)

\bibitem{10.1145/3373719}
Meyer, J., Kay, J., Epstein, D.A., Eslambolchilar, P., Tang, L.M.: A life of
  data: Characteristics and challenges of very long term self-tracking for
  health and wellness. ACM Trans. Comput. Healthcare  \textbf{1}(2) (Mar 2020).
  \doi{10.1145/3373719}

\bibitem{SHELGIKAR2016732}
Shelgikar, A.V., Anderson, P.F., Stephens, M.R.: Sleep tracking, wearable
  technology, and opportunities for research and clinical care. Chest
  \textbf{150}(3),  732 -- 743 (2016)

\bibitem{signal2017children}
Signal, L., Stanley, J., Smith, M., Barr, M., Chambers, T., Zhou, J., Duane,
  A., Gurrin, C., Smeaton, A., McKerchar, C., et~al.: Children's everyday
  exposure to food marketing: an objective analysis using wearable cameras. The
  International Journal of Behavioral Nutrition and Physical Activity
  \textbf{14}(1),  137--137 (2017)

\bibitem{tuovinen2019remote}
Tuovinen, L., Smeaton, A.F.: Remote collaborative knowledge discovery for
  better understanding of self-tracking data. In: 2019 25th Conference of Open
  Innovations Association (FRUCT). pp. 324--332. IEEE (2019)

\bibitem{vizer2009detecting}
Vizer, L.M.: Detecting cognitive and physical stress through typing behavior.
  In: CHI ’09 Extended Abstracts on Human Factors in Computing Systems. p.
  3113–3116. CHI EA ’09, Association for Computing Machinery, New York, NY,
  USA (2009)

\bibitem{WANG2018249}
Wang, P., Sun, L., Smeaton, A.F., Gurrin, C., Yang, S.: {Chapter 9 - Computer
  Vision for Lifelogging: Characterizing Everyday Activities Based on Visual
  Semantics}. In: Leo, M., Farinella, G.M. (eds.) Computer Vision for Assistive
  Healthcare, pp. 249 -- 282. Computer Vision and Pattern Recognition, Academic
  Press (2018)

\end{thebibliography}


\begin{thebibliography}{8}
\bibitem{ref_article1}
Author, F.: Article title. Journal \textbf{2}(5), 99--110 (2016)

\bibitem{ref_lncs1}
Author, F., Author, S.: Title of a proceedings paper. In: Editor,
F., Editor, S. (eds.) CONFERENCE 2016, LNCS, vol. 9999, pp. 1--13.
Springer, Heidelberg (2016). \doi{10.10007/1234567890}

\bibitem{ref_book1}
Author, F., Author, S., Author, T.: Book title. 2nd edn. Publisher,
Location (1999)

\bibitem{ref_proc1}
Author, A.-B.: Contribution title. In: 9th International Proceedings
on Proceedings, pp. 1--2. Publisher, Location (2010)

\bibitem{ref_url1}
LNCS Homepage, \url{http://www.springer.com/lncs}. Last accessed 4
Oct 2017
\end{thebibliography}
%

\end{document}